\documentclass[conference]{IEEEtran}

\usepackage{amsmath,amssymb,latexsym,hhline,graphicx,enumerate,cite}
\usepackage{url}
\usepackage{float}
\DeclareGraphicsExtensions{.eps}
\usepackage{mathtools}
\usepackage{epstopdf}

\DeclarePairedDelimiter\ceil{\lceil}{\rceil}
\DeclarePairedDelimiter\floor{\lfloor}{\rfloor}

\makeatletter

\def\ps@headings{%
\def\@oddhead{\mbox{}\scriptsize\rightmark \hfil \thepage}%
\def\@evenhead{\scriptsize\thepage \hfil \leftmark\mbox{}}%
\def\@oddfoot{}%
\def\@evenfoot{}}
\makeatother
\renewcommand{\cal}{\mathcal}
\newcommand{\nchoosek}[2]{\binom{{#1}}{{#2}}}
\newcommand{\Ps}{\mathbf{p}_{\mathrm{s}}}

\newcommand{\Ibm}{\bm{\mathrm{I}}}

\newcommand{\mA}{\mathcal{A}}

\usepackage{etoolbox}

\patchcmd{\IEEEproofindentspace}{2\parindent}{0pt}{}{}

 \usepackage[normalem]{ulem}
 \usepackage{amsthm}
 \newtheoremstyle{mystyle}
 {2mm}
 {2mm}
 {\itshape}
 {}
 {}
 {.}
 {2mm }
 {}
 \theoremstyle{mystyle}
 
 \newtheorem{ex}{Example}
 \newtheorem{theo}{Theorem}

 \newtheorem{Def}{Definition}

\IEEEoverridecommandlockouts

\usepackage{bm}
\usepackage{color}
\usepackage{algorithm}

\usepackage{cases}

\usepackage{algpseudocode}

\makeatletter

\def\ps@headings{%
	\def\@oddhead{\mbox{}\scriptsize\rightmark \hfil \thepage}%
	\def\@evenhead{\scriptsize\thepage \hfil \leftmark\mbox{}}%
	\def\@oddfoot{}%
	\def\@evenfoot{}}
\makeatother
\begin{document}

\title{Storage Allocation for Multi-Class Distributed Data Storage Systems}

\author{
   Koosha Pourtahmasi Roshandeh, Moslem Noori,~\IEEEmembership{Member,~IEEE,}, Masoud~Ardakani,~\IEEEmembership{Senior~Member,~IEEE,}\\
    and~Chintha~Tellambura,~\IEEEmembership{Fellow,~IEEE}\\
   Department of Electrical and Computer Engineering, University of Alberta, Edmonton, AB, Canada\\
    Email: \{pourtahm, moslem, ardakani, ct4 \}@ualberta.ca\
       }

\maketitle

\begin{abstract}
Distributed storage systems (DSSs) provide a scalable solution for reliably storing massive amounts of data coming from various sources. Heterogeneity of these data sources often means different data classes (types) exist in a DSS, each needing a different level of quality of service (QoS). As a result, efficient data storage and retrieval processes that satisfy various QoS requirements are needed. This paper studies storage allocation, meaning how data of different classes must be spread over the set of storage nodes of a DSS. More specifically, assuming a probabilistic access to the storage nodes, we aim at maximizing the weighted sum of the probability of successful data recovery of data classes, when for each class a minimum QoS (probability of successful recovery) is guaranteed. Solving this optimization problem for a general setup is intractable. Thus, we find the optimal storage allocation when the data of each class is spread minimally over the storage nodes, i.e. minimal spreading allocation (MSA). Using upper bounds on the performance of the optimal storage allocation, we show that the optimal MSA allocation approaches the optimal performance in many practical cases. Computer simulations are also presented to better illustrate the results.


\end{abstract}

\begin{IEEEkeywords}
Distributed storage systems, storage allocation, minimal spreading allocation, multi-class.
\end{IEEEkeywords}

\section{Introduction}

Distributed storage systems (DSSs) play a vital role in numerous innovative and cost-effective services by providing reliable anytime/anywhere access to immense amount of data. For this, different types of data are stored with redundancy over a networked set of storage nodes. That is, for storing a data chunk, the DSS server (controller) first encodes it according to an error-correction coding scheme. Then, the encoded data is partitioned into multiple pieces and each piece is stored over a storage node. 

When a request to download the data is submitted to the server, it tries to access all or a subset of the storage nodes in order to obtain pieces of the encoded data. If the server receives enough pieces of the encoded data, the original chunk can be successfully recovered to serve the download request. However, due to hardware/software failure or network congestion, the server may not receive enough pieces of the encoded data resulting in data recovery failure. To quantitatively account for these circumstances, the probability of successful data recovery by the server, denoted by $P_{\mathrm{s}}$, is commonly used in the literature \cite{leong2012distributed, Sardari_Allocation_2010, ntranos2012allocations}.

The way that the server partitions and stores the coded data over the storage nodes is often referred to as storage or memory allocation \cite{leong2012distributed, Sardari_Allocation_2010}. It is known that storage allocation affects different performance measures of a DSS such as $P_{\mathrm{s}}$ \cite{ Sardari_Allocation_2010, hong2016asymptotic, ntranos2012allocations, Li2013Allocation, Noori2015k-guaranteed, andriyanova2015distributed}, service rate \cite{LeongOptimaldelay2011, joshi2014delay, Noori16Servicerate} as well as storage and repair cost \cite{yu2011minimization}. Hence, several studies have been dedicated to improving the performance of DSSs via careful storage allocation. In this work, we focus on studying the effect of storage allocation on the successful data recovery, $P_{\mathrm{s}}$. 
 
Finding the optimal storage allocation to maximize $P_{\mathrm{s}}$ is known to be a quite challenging problem and the optimal storage allocation for a general DSS setup has yet to be found \cite{leong2012distributed}. Nevertheless, several studies have addressed maximum-$P_{\mathrm{s}}$ storage allocation for specific setups and under various assumptions. For instance, optimal quasi-uniform allocation for maximizing $P_{\mathrm{s}}$, where a fixed-size randomly chosen subset of storage nodes is accessed by the server, is studied in \cite{Sardari_Allocation_2010}. Also, two approximation algorithms are proposed in \cite{ntranos2012allocations} to maximize $P_{\mathrm{s}}$ in a heterogeneous DSS where storage nodes have different reliabilities. Moreover, for a DSS with heterogeneous nodes in terms of storage capacity, an iterative algorithm has been proposed in \cite{Noori2015k-guaranteed} to find a $k$-guaranteed allocation\footnote{An allocation is said to be $k$-guaranteed if the stored data can be recovered by accessing any arbitrary set of $k$ storage nodes}. In another study \cite{Noori16Servicerate}, it is shown that considering the effect of successful data recovery on the service rate and assuming exponential waiting time at the storage nodes, the service rate is maximized through uncoded data replication over the storage nodes. 
 
While the aforementioned studies have shed some light on the optimal storage allocation, they fall short of addressing a practical aspect of DSSs, that is the heterogeneity of the stored data. To be more specific, a DSS usually stores different classes (types) of data often coming from different sources \cite{Kumar_Multi_2015}. We call such a DSS a \emph{multi-class} DSS where each class requires its own level of quality of service (QoS). For instance, Amazon S3 allows its customers to choose from three storage classes offering different levels of durability, reliability, and availability \cite{Amazon}. While such asymmetric QoS requirements has to be taken into account in the storage allocation, to the best of our knowledge, it has not been considered in the previous studies. Note that in a multi-class DSS, storage allocation for different classes are intertwined due to the limit on the available storage space. This makes storage allocation for a multi-class DSS a more challenging problem compared to the storage allocation for a single-class DSS.

In this paper, we study the problem of storage allocation for a multi-class DSS. More specifically, we consider storing $K$ classes of data over a DSS with $N$ storage nodes. To account for possible access failures (e.g., due to network congestion), we adopt the probabilistic access model \cite{leong2012distributed} where each storage node fails to respond to the server's access request with a given probability $q$. Assuming a storage budget $T_i$ for the $i^{\text{th}}$ class of data, we focus on maximizing the weighted sum of the probability of successful recovery of all data classes where weights reflect the QoS requirement of each class. Further, to guarantee a minimum viable service for each class, a constraint on the minimum probability of successful recovery for each class is introduced in the optimization problem. 

The analytical solution of this optimization problem is, however, intractable even if there is only one class in the network, i.e. a single-class DSS. That said, we narrow down our attention to the optimal minimal spreading allocations (MSA) to maximize the considered weighted sum. It is worth mentioning that MSA is the optimal symmetric allocation for maximizing $P_{\mathrm{s}}$ in some single-class DSS setups \cite{leong2012distributed} motivating us to contemplate its performance for the considered multi-class DSSs. Furthermore, MSA is the optimal allocation in terms of maximizing the service rate in a single-class DSS with exponential waiting time at the users \cite{Noori16Servicerate}. In addition, another study \cite{LeongOptimaldelay2011} shows that MSA minimizes the expected recovery delay when the storage budget is an integer.

Assuming MSA as the storage allocation, we first formulate finding optimal MSA to maximize the weighted sum of the probabilities of successful recovery as a non-linear integer optimization. Then, an iterative algorithm with time complexity $O(N)$ is presented to solve this optimization problem. Despite the optimality of this iterative solution, its linear complexity may become a challenge when applied to large-scale DSSs. To address this issue, we also present a suboptimal solution for the considered weighted-sum maximization problem. This suboptimal solution has a worst-case complexity of $O(K)$ and the resulting MSA either matches or slightly underperforms the optimal MSA. In the next step,  the performance of the presented (sub)optimal MSA is compared with the upper bound on the weighted sum of the probabilities of successful recovery when no assumption on the format of the allocation policy is made. We analytically prove that optimal MSA achieves this upper bound for a significant range of $q$ meaning that MSA is indeed the optimal allocation in those ranges.

\section{System Model}
In this section, we explain the data storage and server access model in the considered multi-class DSS.

\subsection{Storage model}

\begin{figure}%
\vspace{-1.7 cm}
	\includegraphics[width=\columnwidth]{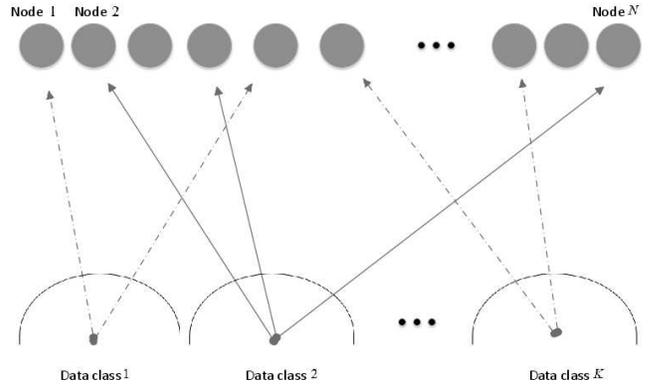}%
	\caption{System Model.}%
	\label{Fig: System Model}%
\end{figure}

Here, we consider the storage of $K$ classes of data over a DSS. More specifically, $k$ data blocks from each of the $K$ classes of data are to be stored over a DSS with $N$ equal-capacity storage nodes. For clarity of presentation, let us assume that the storage capacity of each node is also $k$ blocks.  Later in Section \ref{Sec: discussion}, we explain how this assumption can be relaxed to apply our results to more general cases. The storage size for class $i  \in \mathcal{K} = \{1,2,\ldots,K\}$ is limited by a storage budget $T_i$ which can reflect the QoS of class $i$. 

To store the data of class $i \in \mathcal{K} = \{1,2,\ldots,K\}$, its $k$ blocks of data are first encoded by a suitable minimum distance separable (MDS) code \cite{leong2012distributed, ntranos2012allocations} to form $n_i$ coded blocks such that $n_i \leq T_i$. Having these $n_i$ encoded data blocks, the server then spreads them over the $N$ storage nodes according to a \emph{storage allocation}  policy formally defined later in this section. That said, the server can retrieve the $k$ data blocks of class $i$ if it successfully receives at least $k$ out of the $n_i$ coded blocks from the storage nodes when a download request is submitted. A graphical illustration of the system model is presented in Fig.~\ref{Fig: System Model}.

For simplicity of presentation, we normalize all storage capacities and data sizes with $k$. That is, the normalized capacity of each storage node $n$, denoted by $c_n$, is equal to one for $ n \in \mathcal{N}=\{1,2,...,N\}$. Further, $0 \leq x_{i,n} \leq c_n = 1$ denotes the normalized number of encoded blocks from class $i$ that are stored over storage node $n$. With some abuse of notations, we keep using $n_i$ and $T_i$ for the normalized number of coded blocks and storage budget of class $i$. To ensure that the budget limits and nodes' storage limits are not violated, we have $\sum_{n \in \cal{N}} x_{i,n} = n_i \leq T_i$ and  $\sum_{i \in \cal{K}} x_{i,n} \leq c_n =  1$. In the following, we use a storage budget vector $\cal{T} = (T_1,T_2,\ldots,T_N)$  refer to the storage budget of all classes. 

Now that we have explained the storage model, we formally define the allocation policy.  

\begin{Def}
A storage allocation policy elucidates the way the encoded blocks of the $K$ classes are stored over the $N$ nodes and is identified by a $K$-tuple $\mA=(\mA_1,\mA_2,...,\mA_K)$  where $\mA_i=(x_{i,1}, x_{i,2},..., x_{i,N})$ for all $i \in \cal{K}$.
\end{Def}
Beside determining how coded data blocks are spread over the storage nodes, a storage allocation policy $\cal{A}$ also implicitly identifies the rate of the MDS codes for each data class.

%

\subsection{Access model}
We adopt a probabilistic access model \cite{leong2012distributed} is adopted at the server. That is, the server tries to access all the storage nodes when a request to download the data of a class, say class $i$, is submitted. However, the server's access attempt to a node may fail due to a hardware failure or congestion at the node. Hence, a probability of successful access $p < 1$ is considered. The server is able to recover the data of class $i$ if $\sum_{n \in \bf{r}} x_{i,n} \geq 1 $ where $\bf{r}$ denotes the set of nodes which have been successfully accessed by the server.  Hence, the probability of successful recovery of the class $i$'s data is 
\begin{equation} \label{Eq P_s random}
P_{\mathrm{s},i} = \sum_{\bm{r} \subseteq \cal{N}} p^{\vert \bm{r} \vert} (1 - p)^{N - \vert \bm{r} \vert} \Ibm \left[ \sum_{n \in \bf{r}} x_{i,n} \geq 1 \right].
\end{equation}

\noindent where $\Ibm[G]$ is the indicator function meaning that $\Ibm[G] = 1$ if $G$ is true, and $\Ibm[G] = 0$ otherwise.

Given an allocation policy $\mA=(\mA_1,\mA_2,...,\mA_K)$, the $K$-tuple $\Ps = [P_{\mathrm{s},i}]_{i \in \mathcal{K}}$ is a feasible vector of probabilities of successful recovery if the data of the $i^{\text{th}}$ class can be successfully recovered with probability $P_{\mathrm{s},i}$. We find the following definition useful for our problem definitions in Section \ref{motivforkclasses}.

\begin{Def}
The successful recovery region of a multi-class DSS, denoted by $\mathbf{\Theta}$, is defined as the union of all feasible vectors $\Ps = [P_{\mathrm{s},i}]_{i \in \mathcal{K}}$ given a storage budgets vector $\cal{T}$. 
\end{Def}

\section{Motivation and Problem Definition} \label{motivforkclasses}
The effect of the storage allocation policy on the  probability of successful data recovery has been investigated in several prior studies for single-class DSSs. To this end, optimal storage allocation policies that maximize the probability of successful recovery in single-class DSSs have been proposed for various setups. For more details, the interested reader is referred to \cite{leong2012distributed,Sardari_Allocation_2010,ntranos2012allocations}. 

Finding such optimal allocation policies, however, is more challenging for a multi-class DSS as the allocation policy of each class potentially affects the probability of successful recovery of other classes. Further, in a multi-class DSS, the allocation policy should reflect the reliability needs, measured by the probability of successful recovery, of each of the diverse data classes. 

Here, to improve the overall performance of the system  while taking into account different classes' service requirements, we consider optimizing a weighted sum of $P_{\mathrm{s},i}$'s, where a higher weight is assigned to more important data classes. We aim at solving this optimization problem under the constraint that for any class $i$, $P_{\mathrm{s},i} \geq P_{\mathrm{s},i}^{\mathrm{min}}$, to ensure a minimum viable service for it. Denoting the weight of class $i$ by $\alpha_i > 0$, the weighted sum maximization problem is formulated as:
\begin{equation*}\tag{P1}
\begin{aligned}\label{P1} 
& \underset{\cal{A}}{\text{maximize}}
&&  \langle \boldsymbol{\alpha}, \Ps \rangle \\
& \text{subject to}
&& \Ps \in \mathbf{\Theta}, \\
&&& P_{\mathrm{s},i}^{\mathrm{min}} \leq P_{\mathrm{s},i} \, \, \, \forall i \in \cal{K}, 
\end{aligned}
\end{equation*}
where $\boldsymbol{\alpha}=[\alpha_i]_{i \in \cal{K}}$ is a vector containing the weights associated with the classes and $\langle \cdot,\cdot \rangle$ is the inner product of the vectors.

The difficulty of solving (\ref{P1}) lies in finding  $\mathbf{\Theta}$, otherwise the objective function and the constraint on minimum $P_{\mathrm{s},i}$ are simple linear functions. An analytical description of  $\mathbf{\Theta}$, however, is intractable due to the strong interdependency of the allocation policy of one class and the probability of successful recovery of other classes. The following example helps explaining such an interdependency for a simple DSS.

\begin{ex} \label{Example 1} 
	Consider a DSS with $N = 3$ nodes where $c_n = 1,~ \forall n \in \mathcal{N}$ and $\cal{T} = (\frac{3}{2},\frac{5}{4})$. Table~\ref{Tab Example} shows four possible allocations for this setup alongside their corresponding probability of successful recovery. As seen from this table, if $\alpha_1 \gg \alpha_2 $, then Case 1 and Case 4 are optimal for $p \leq \frac{1}{2}$ and $\frac{1}{2} < p$ respectively resulting in the maximum $P_{\mathrm{s},1}$. On the other hand, for $\alpha_1 \ll \alpha_2$ or $\alpha_1 = \alpha_2$, only Case 1 results in the optimal solution of (\ref{P1}). 
\hfill $\blacksquare$
\end{ex}

\begin{table}[ht!]
	\begin{center} 
		\caption{Different allocations and their corresponding probability of successful recovery.}
		\begin{tabular}{|c|c|c|}
			\cline{1-3}
			Allocation & $P_{\mathrm{s},1}$ & $P_{\mathrm{s},2}$ \\ \hline
			Case 1: $\mA_{1} = (1,\frac{1}{2}, 0)$, $\mA_{2} = (0,\frac{1}{4}, 1)$& $p$ & $p$ \\ \hline
			Case 2: $\mA_{1} = (1,\frac{3}{8}, \frac{1}{8})$, $\mA_{2} = (0,\frac{5}{8}, \frac{5}{8})$& $p$ & $p^2$ \\ \hline
			Case 3:$\mA_{1} = (\frac{3}{4},\frac{2}{4}, \frac{1}{4})$, $\mA_{2} = (\frac{1}{4},\frac{1}{4},\frac{3}{4})$& $2p^2 - p^3$ & $2p^2 - p^3$ \\ \hline
			Case 4: $\mA_{1} = (\frac{1}{2},\frac{1}{2}, \frac{1}{2})$, $\mA_{2} = (\frac{5}{12},\frac{5}{12},\frac{5}{12})$& $3p^2 - 2 p^3$ & $p^3$ \\ \hline
		\end{tabular}
		\label{Tab Example}
	\end{center}
\end{table}

Since (P1) cannot be solved in a general setup, here we focus on a special, yet practically important, case of the storage allocation, called minimum spreading allocation (MSA), formally defined in the following. 

\begin{Def} \label{Def MSA} 
In a minimal spreading allocation policy, for each class $i$, we have	
	\[
	x_{i,n} =
	\begin{cases} 
	\hfill 1    \hfill & \forall n \in \mathcal{V}_i  \\
	\hfill 0 \hfill & \text{ otherwise} \\
	\end{cases}
	\]	
where $\mathcal{V}_i \subset \mathcal{N}$ is the set of nodes storing the data of class $i$. 
\end{Def}

It has been shown that MSA is optimal in terms of expected recovery delay, average service rate and probability of successful recovery for several setups of single-class DSSs \cite{leong2012distributed, Noori16Servicerate, LeongOptimaldelay2011}. Moreover, by adopting an MSA policy, the data of each class could be stored through replication \cite{leong2012distributed} removing the need for devising custom-tailored storage coding schemes. That said, we focus on finding the optimal MSA for the considered multi-class DSS in the rest of the paper. 

From Definition~\ref{Def MSA}, one can see that an MSA allocation is fully identified by $\vert \cal{V}_i \vert$, denoted by $x_i$ from now on, as $x_{i,n}$ is either 0 or 1. 

Notice than when $\sum_{i=1}^{K} \floor{T_i} \leq N$, the MSA optimization has a trivial solution of $x_i = \floor{T_i}, ~ \forall i \in \mathcal{K}$. Hence, we only consider $\sum_{i=1}^{K} \floor{T_i} > N$ in the sequel. Furthermore, $P_{\mathrm{s},i}=1-q^{x_i}$ where $q=1-p$. This is because  the recovery of the class $i$ data fails when all $x_i$ nodes containing its data fail. Hence, finding the optimal MSA to maximize the weighted sum can be formulated as the following minimization problem
\begin{equation*}\tag{P2}
\begin{aligned} \label{P2}
& \underset{\{x_i \}_{i \in \cal{K}}}{\text{minimize}}  &   &\sum_{i=1}^{K} \alpha_i q^{x_i}  \\
& \text{subject to}                         &   &\sum_{i=1}^{K} x_i \leq N, \\
&                                           &   & \sum_{i=1}^{K} \floor{T_i} > N, \\
&                                           &   & x_i^{\mathrm{min}} \leq x_i \leq \floor{T_i}  ~~~\forall i \in \mathcal{K}, \\
\end{aligned}
\end{equation*}
where $x_i \in \mathbb{Z^+}$  and $x_{i}^{\mathrm{min}}= \ceil {\log_q (1-P^{\mathrm{min}}_{\mathrm{s},i} ) }$. Note that in the above optimization problem, the first constraint makes sure that the total available storage capacity is not exceeded. Further, the minimum requirement on the probability of successful recovery and budget limit of each class are enforced through the last constraint\footnote{Here, it is assumed that $x_{i}^{\mathrm{min}} \leq T_i$ and $\sum_{i \in \cal{K}} x_i^{\mathrm{min}} \leq N$ as the problem is infeasible otherwise.}. 

Now, with a change of variable $y_i = x_i - x_i^{\mathrm{min}}$, the optimization problem in (\ref{P2}) is transformed into the following
\begin{equation*}\tag{P3}
\begin{aligned} \label{P3}
& \underset{\{y_i \}_{i \in \cal{K}}}{\text{minimize}}  &   &\sum_{i=1}^{K} \beta_i q^{y_i}  \\
& \text{subject to}                         &   &\sum_{i=1}^{K} y_i \leq N', \\
&                                           &   & \sum_{i=1}^{K} \floor{T'_i} > N', \\
&                                           &   & 0 \leq y_i \leq \floor{T'_i}  ~~~\forall i \in \mathcal{K}, \\
\end{aligned}
\end{equation*}
where $\beta_i = \alpha_i q^{x_i^{\mathrm{min}}}$, $N' = N - \sum_{i \in \cal{K}} x_i^{\mathrm{min}}$ and $T'_i = T_i - x_i^{\mathrm{min}}$.

Note that here, (\ref{P3}) resembles (\ref{P2}) $x_i^{\mathrm{min}} = 0$, $\forall i \in \cal{K}$. Hence, for the ease of notations and discussions, in the rest of this work, we focus on (\ref{P2}) were all $x_i^{\mathrm{min}}$'s are set to zero. The optimization problem in (\ref{P2}) is a non-linear integer optimization problem. In the following section, we present our main results on how the storage allocation problem in (\ref{P2}) can be tackled.

\section{Main results} \label{Sec: main results}

\subsection{Exact solution of \eqref{P2}} \label{subsec:optimal solution}
The exact solution of \eqref{P2} can be found using an iterative approach where we assign the available storage units one at the time. 

Imagine that we want to assign the first unit of storage. It is clear that, in order to minimize $\sum \alpha_i q^{x_i},$ this unit must be assigned to the class with the largest $\alpha_i$, let us call this class $j$. Now, the storage allocation problem can be updated to one where the total storage budget is reduced to $N-1$ and class $j$ already has one storage unit assigned to it. Hence, if the optimal solution for class $j$ is $x_j$ in the original problem, in the updated problem it is $x_j-1.$ Also, note that class $j$ contributes to $\sum \alpha_i q^{x_i}$ as $(\alpha_j q) q^{x_j-1}$. Therefore, the updated problem is similar to the original with two minor differences: $N$ is updated to $N-1$ and $\alpha_j$ is updated $\alpha_j q.$ That is, the second unit of storage can now be assigned following similar steps. Repeating this approach, we can find the optimal solutions. 

The only other thing that we have to be careful about is that we should not let any data class to violate its storage limit. For this, if a class reaches its limit, we set $x_i$ for this class to $\lfloor T_i \rfloor,$ we remove this class form the problem and we continue. This recursive procedure is continued until all $N$ storage units are assigned. The above procedure is presented as Algorithm~\ref{alg:optimal}. 

\begin{algorithm}[t]
	\caption{}\label{alg:optimal}
	\begin{algorithmic}[1]
		\Procedure{}{$N, \mathcal{K},\cal{T}, \boldsymbol{\alpha}$}

		\State $x^{opt}_i \leftarrow 0  ,~ \forall i \in \mathcal{K} $
		\While {$N>0$}
		
		\State $\alpha_{j} \leftarrow max(\boldsymbol{\alpha}) $
		\State $ x^{opt}_j \leftarrow x^{opt}_j+1$
		\State $\alpha_j  \leftarrow q \alpha_j$
		\State	$N \leftarrow N-1$

		\If	{$x_j = \floor{T_j} $ }		
		\State		$\boldsymbol{\alpha} \leftarrow ( \alpha_1, ..., \alpha_{j-1}, \alpha_{j+1}, ..., \alpha_K  ) $		
		\EndIf	
		
		\EndWhile

		\EndProcedure
	\end{algorithmic}
\end{algorithm}

The time complexity of this solution is $O(N).$ This is because, we assign the available storage units one at a time. Hence, we end up repeating the assignment process $N$ times. After every assignment, we need to perform one comparison to find the new largest weight. When $N$ is large, this complexity order may not be acceptable. Hence, in the next section, we suggest another approach for solving \eqref{P2} that in many cases gives the optimal solution. If not, the solution is close to optimal. The complexity order of the new algorithm is $O(K)$ which can be much smaller than $O(N)$. Moreover, since in some cases the solution of the new approach is in closed form, the relation between system parameters and the optimal solution is better seen compared to the iterative approach presented above.

\subsection{A low-complexity near-optimal solution of (P2)} \label{subsec:suboptimal solution}

The optimization problem \eqref{P2} can be directly solved if we remove the budget constraints for data classes (or equivalently $T_i=N ,~ \forall i \in \mathcal{K}$) and let $x_i \in \mathbb{R} ,~ \forall i \in \mathcal{K}$. The following lemma gives the optimal solution of this relaxed version of \eqref{P2}.

\begin{theo}
	\label{easy sol}
	Consider  $x_i \in \mathbb{R} ,~ \forall i \in \mathcal{K}$. Then, $\forall i, x_i=\frac{N}{K} + \frac{1}{K} \log_q \frac{\prod_{j=1, j\neq i }^{K} \alpha_j}{\alpha_i ^ {K-1}  }$ minimizes $\sum_{i=1}^K \alpha_i q^{x_i}$ subject to $\sum_{i=1}^K {x_i} \le N$.
\end{theo}

\begin{IEEEproof}
	Using geometric-mean arithmetic-mean inequality, we have
	\begin{equation} \label{ineq zero state}
	\begin{aligned}
	&\sum_{i=1}^{K}  q^{x_i + \log_q \alpha_i} &  &\geq  K \sqrt[ K]{\prod_{i=1}^K q^{x_i + \log_q \alpha_i}}\\
	&                                         &  &= K q^{ \frac{\sum_{i=1}^{K} x_i   +\log_q \prod_{ i=1 }^{K} \alpha_i}{K} }
	\end{aligned}
	\end{equation} 
	
	\vspace{-0.7em}
	\noindent Now, notice that the right-hand side of \eqref{ineq zero state} is minimized when $\sum_{i=1}^{K} x_i=N$.	
	Moreover, arithmetic-mean achieves its lower bound when 
	\vspace{+0.5em} 
	\begin{eqnarray}\label{equality zero state}
	\begin{aligned} 
	& \forall i,j \in \mathcal{K} ~,~q^{x_i + \log_q \alpha_i}  &   &=q^{x_j + \log_q \alpha_j}  \\
	&                            &   &= q^{ \frac{N   +\log_q \prod_{ i=1 }^{K} \alpha_i}{K}}  \\
	&                            &   &= D    \\
	\end{aligned}
	\end{eqnarray}
	
	\vspace{+0.5em}
	\noindent where $D$ is a constant. Solving \eqref{equality zero state} for $x_i$ results in $x_i=\frac{N}{K} + \frac{1}{K} \log_q \frac{\prod_{j=1, j\neq i }^{K} \alpha_j}{\alpha_i ^ {K-1}  }$. 	
\end{IEEEproof}

While Lemma \ref{easy sol} provides the optimal real-valued solutions for the relaxed version of \eqref{P2} with no individual budget constraints, the actual solutions of \eqref{P2} are non-negative integers in the range of the budget constraints of different data classes.  

In the following we show that it is possible to use Lemma~\ref{easy sol} as a baseline to form a low-complexity and near-optimal solution for \eqref{P2}. The overview of our approach is that we use the unconstrained values obtained in Lemma~\ref{easy sol} to partition the set of $K$ data classes into three subsets.  Based on the cardinality of these subsets, three different cases are identified. In Case 1, we directly propose the optimal solutions of  \eqref{P2}. In Case 2, we provide an iterative approach with worst-case complexity order $O(K)$ for solving \eqref{P2}. In Case 3, we propose a sub-optimal solution of   \eqref{P2}, again with worst-case complexity $O(K).$ Our numerical results in Section \ref{Sec: simulation results} show that the performance of this solution is very close to the optimal solution of Section \ref{subsec:optimal solution}. Since typically $K \ll O$, the new algorithm is much more efficient than the optimal solution.

Based on the values of $x_i=\frac{N}{K} + \frac{1}{K} \log_q \frac{\prod_{j=1, j\neq i }^{K} \alpha_j}{\alpha_i ^ {K-1}  }$, we partition the set of data classes $\cal K$ into three distinct subsets of $\mathcal{K}_1$, $\mathcal{K}_2$ and $\mathcal{K}_3$ such that that $  \mathcal{K}_1= \{i~ | ~ x_i < 0, i \in \mathcal{K} \} $, $  \mathcal{K}_2= \{i ~| ~ 0 \leq x_i < \floor{T_i}, i \in \mathcal{K} \} $ and $  \mathcal{K}_3= \{i ~|~  x_i \geq \floor{T_i}, i \in \mathcal{K} \} $.  Now, three different cases are possible:

\begin{enumerate}
	\item $|\mathcal{K}_1| = |\mathcal{K}_3| = 0.$   
	\item $|\mathcal{K}_1| = 0$, $|\mathcal{K}_3| \neq 0.$
	\item $|\mathcal{K}_1| \neq 0.$ 
\end{enumerate}
We address these three cases separately.

\subsubsection{Case 1 ($|\mathcal{K}_1| = |\mathcal{K}_3| = 0$)}

In this case, all $x_i$'s obtained using Lemma \eqref{easy sol} satisfy the budget constraints of their respective data classes. The only problem is that these $x_i$'s are not necessarily integers. In two steps, we find the optimal integer-valued solutions. First, using the following theorem we show that the optimal value in class $i$ is either $\lfloor x_i \rfloor$ or $\lceil x_i \rceil$, where $x_i$ is found from Lemma~\ref{easy sol}. Then, in another theorem, we identify which classes should use the ceiling and which classes the floor of their corresponding $x_i$.

\begin{theo}
	\label{theorem state 2}
	For $x_i$ resulted from Lemma~\ref{easy sol}, assume $\exists i \in \mathcal{K}, x_i \notin \mathbb{Z}$. Also, assume $|\mathcal{K}_1| = |\mathcal{K}_3| = 0$. Then, the optimal solution of problem  \eqref{P2} in class $i$ is either
	$ x^{opt}_i= \floor{x_i}$ or $ \floor{x_i}+1 ,~\forall i \in \mathcal{K}$.
\end{theo}

\begin{IEEEproof}
	See Appendix \ref{Appendix G}.
\end{IEEEproof}

\begin{theo}
	\label{Theo state 2}
	In Theorem~\ref{theorem state 2}, let  $ x_i= \floor{x_i}+ e_i ,~\forall i \in \mathcal{K}$. Without loss of generality, assume  $e_1 \geq e_2 \geq ... \geq e_K$, and  \mbox{$ N-\sum_{i=1}^{K}\floor{x_i} =M$}. Then, the optimal solution of problem  \eqref{P2} can be obtained as
	
	\begin{equation}x^{opt}_i=	
	\begin{cases}{}
	\floor{x_i}+1 &    1 \leq i \leq M \nonumber
	\\
	\floor{x_i} &  M < i \leq K  \nonumber
	\end{cases}	
	\end{equation}
\end{theo}

\begin{IEEEproof}
	See Appendix \ref{Appendix H}.	
\end{IEEEproof}
Simply put, Theorem~\ref{Theo state 2} states that the $M$ classes with the largest non-integer part must receive $\lceil x_i \rceil$ and the other classes must receive $\lfloor x_i \rfloor$ storage units. Notice that in this case, we directly derived the optimal solution of \eqref{P2}.

\subsubsection{Case 2 ($|\mathcal{K}_1| =0$, $|\mathcal{K}_3| \neq 0$)}

Here, $x_i>0 ,~ \forall i \in \mathcal{K}$, but data classes that belong to $\mathcal{K}_3$ have $x_i$'s that are greater than their corresponding budget limit. The next Theorem gives the optimal solution of data classes in $\mathcal{K}_3$.

\begin{theo}
	\label{Theo case 2}
	Assume $|\mathcal{K}_1| = 0$. Then, the optimal solution of problem \eqref{P2} has the property $x^{opt}_i= \floor{T_i} ,~ \forall i \in \mathcal{K}_3$.
\end{theo}

\begin{IEEEproof}	
	By contradiction and following similar steps to the proof of Theorem \ref{theorem state 2}, one can easily prove this theorem. A detailed proof is therefore omitted.	
\end{IEEEproof}

Using Theorem \ref{Theo case 2}, we have the optimal solution of optimization problem \eqref{P2} for data classes in $\mathcal{K}_3 $. Now, we can repeat solving \eqref{P2} with the updated values $\mathcal{K}^{new}= \mathcal{K} \setminus \mathcal{K}_3$ and $N^{new}=N-\sum_{i \in \mathcal{K}_3 }^{} \floor{T_i}$.

The new optimization problem has fewer data classes and may end up being a problem identified by Case 1, 2 or even 3 (discussed below). Nonetheless, as the problem size is smaller, even in the worst-case, after a maximum of $K$ rounds, \eqref{P2} is solved.

\subsubsection{Case 3 ($|\mathcal{K}_1| \neq 0$)}

Here, since $\mathcal{K}_1 \neq \emptyset$, there are some negative values among $x_i$'s resulted from Lemma~\ref{easy sol}. Moreover, we may have some values that go beyond the budget constraint of data classes (depending on the cardinality of $\mathcal{K}_3$). The complexity of this case is rooted in many different scenarios that can happen. To see this complexity, note that if we force $x_i^{opt}=0$ for a class with $x_i<0$ and update $x_i$'s from Lemma~\ref{easy sol}, some of the classes that used to be in $\mathcal{K}_3$ may no longer be there. Likewise, if we force $x_i= \floor{T_i}$ for some of the members of $\mathcal{K}_3$, the elements of $\mathcal{K}_1$ will change. Also, the order that we choose to set $x_i$'s to zero or $\floor{T_i}$ affects the final solution and our study suggests that the optimal ordering (even if possible to identify) does not follow any easy procedure. Hence, to fulfill our main goal of having a low-complexity solution, we suggest the following approach, which later is numerically tested and gives great results. 

After using Lemma~\ref{easy sol} to find $x_i$'s, we let $ x^{opt}_i=0 ,~ i \in \mathcal{K}_1$. Now, we can repeat solving \eqref{P2} with the updated value $\mathcal{K}^{new}= \mathcal{K} \setminus \mathcal{K}_1$. Again, note that the new problem has fewer classes, and that it may be a problem in any of the above three cases (most likely in Case 1). Therefore, in Case 3, similar to Case 2, the worst-case complexity is $O(K).$

Please note that in cases 1 and 2, we do not deviate from the actual optimal solutions, and only in Case 3 we may lose optimality. Moreover, Case 3 is a rare one because it needs some very low priority classes (classes with very small weight compared to others) such that Lemma~\ref{easy sol} gives rise to negative $x_i$'s for them. Even if this is the case, assigning zero storage to those classes should not affect the overall performance much, as they had very small weights compared to others. Hence, we expect this procedure to perform well. This is indeed verified by our numerical results in Section \ref{Sec: simulation results}.   

We have summarized the above approach to sub-optimally solve \eqref{P2} in  Algorithm \ref{alg:WOMSA}.

\begin{algorithm}[t]
	\caption{}\label{alg:WOMSA}
	\begin{algorithmic}[1]
		\Procedure{}{$N, \mathcal{K},\cal{T}, \boldsymbol{\alpha}$}
		\vspace{0.5em}
		
		\While {$|\mathcal{K}| \neq \emptyset$}
		\State $\forall i \in \mathcal{K}$,  $x_i \leftarrow \frac{N}{K} + \frac{1}{K} \log_q \frac{\prod_{j=1, j\neq i }^{K} \alpha_j}{\alpha_i ^ {K-1}}$
		
		\State $  \mathcal{K}_1 \leftarrow \{i |  x_i < 0, i \in \mathcal{K} \} $  
		\State $  \mathcal{K}_2 \leftarrow \{i |  0 \leq x_i < \floor{T_i}, i \in \mathcal{K} \} $
		\State $  \mathcal{K}_3 \leftarrow \{i |  x_i \geq \floor{T_i}, i \in \mathcal{K} \} $

		\If {$|\mathcal{K}_1| = \emptyset$ \& $|\mathcal{K}_3| = \emptyset$ }
		
		\State $M \leftarrow N-\sum_{i=1}^{K} \floor{x_i}$ 	
		\State $e_i \leftarrow x_i- \floor{x_i}$				
		\State For the $M$ greatest values of $e_i$'s 
		\State $x^{opt}_{i} \leftarrow \floor{x_{i}}+1$, otherwise $x^{opt}_i \leftarrow \floor{x_i}$ 
		\State \textbf{Break}	
		
		\ElsIf {$|\mathcal{K}_1| = \emptyset$ \& $|\mathcal{K}_3| \neq \emptyset$}

		\State	$x^{opt}_i  \leftarrow \floor{T_i} ,~ \forall i \in \mathcal{K}_3$
		\State $\mathcal{K} \leftarrow \mathcal{K} \setminus \mathcal{K}_3$
		\State $N \leftarrow N-\sum_{i \in \mathcal{K}_3 }^{} \floor{T_i}$

		\ElsIf {$|\mathcal{K}_1| \neq \emptyset$ }
		
		\State $ x^{opt}_i \leftarrow 0 ,~ \forall i \in \mathcal{K}_1$
		\State $\mathcal{K} \leftarrow \mathcal{K} \setminus \mathcal{K}_1$

		\EndIf

		\EndWhile

		\EndProcedure
	\end{algorithmic}
\end{algorithm}

\subsection{Bounding the performance}

Now that \eqref{P2} is solved (optimally in Section \ref{subsec:optimal solution} or sub-optimally in Section \ref{subsec:suboptimal solution}), it is interesting to study how much we lost by solving MSA instead of solving the general optimal storage allocation formulated in \eqref{P1}. Consider the gap between the weighted sum of the probabilities of successful recovery of different classes for \eqref{P1} and \eqref{P2}. A small gap indicates that instead of solving the highly complex \eqref{P1}, one can resort to solving the much more efficient \eqref{P2} without much performance loss. In other words, a small gap suggests that MSA is close to optimal for our multi-class setup. Hence, we are interested to see when this gap is small. 

As discussed \eqref{P1} is too hard to be solved efficiently. Hence, here we first find an upper bound on the performance of \eqref{P1}. Clearly, the performance gap of MSA and \eqref{P1} is less than or equal to the performance gap between MSA and the upper bound of \eqref{P1}. Also, wherever MSA performs close to this upper bound, we know that MSA is efficient, and there is no need to solve the highly complex \eqref{P1}.

A trivial upper bound on the erformance of \eqref{P2} is obtained by generalizing the results of \cite{leong2012distributed} as 
\begin{eqnarray} \label{eq:bound}
 \sum_{i \in \mathcal{K}}^{} \sum_{r=0}^{n} \alpha_i \min(\frac{r T_i}{N}, 1) \nchoosek{N}{r} p^r (1-p)^{(N-r)}
\end{eqnarray}

 As our numerical results show in Section \ref{Sec: simulation results}, the solution of \eqref{P2} gets very close to this upper bound in a wide range of access probability $p$. This means MSA for multi-class DSS is an efficient solution for those regions. 

In order to analytically identify the region of $p$ where MSA is close to optimal, another simple comparison can be done. We compare the solution of \eqref{P2} with the ideal case where all classes are retrieved perfectly ($P_{\mathrm{s},i}=1 , \forall i \in \mathcal{K}$), where we consider the case that there are no individual budget constraints on the data classes. 

The following theorem identifies the range of access probability $p$ where the performance gap between the optimal MSA and perfect recovery case (described above) is less than $\epsilon.$ This theorem is studied numerically in Section \ref{Sec: simulation results}.

\begin{theo}
	\label{Theorem: gap}
	Assume  $T_i=N ,~ \forall i \in \mathcal{K}$, then the gap between the weighted sum of successful recovery probability of all data classes in optimal MSA and perfect recovery  is less than $\epsilon$ for
\begin{align}	\label{eq theo5}
	 p> 1- \min \{ (\frac{\epsilon^K}{K^K \prod_{i=1}^{K} \alpha_i})^{\frac{1}{N}} ,   (\frac{\alpha^{K-1}_{min}}{\prod_{j=1 , \alpha_j \neq \alpha_{min} }^{K}  \alpha_j} )^{ \frac{1}{|N-1|}  }  \}
\end{align}	 
	  where $\alpha_{min} =  \min\limits_{1\leq i\leq K} \alpha_i$.
\end{theo}

\begin{IEEEproof}
	See Appendix \ref{Appendix J}.
\end{IEEEproof}

According to the proof of Theorem 5, in the proposed range of access probability $p$, our heuristic algorithm gives the optimal MSA.

In the following, we present a generalization of the considered problem. Recall that we assumed that the normalized capacity of all storage nodes is equal to 1, i.e.,  $c_n=1 ,~\forall n \in \mathcal{N}$. The next section discusses situations where $c_n>1$ for some $n$.

\section{Nodes with storage capacity larger than one} \label{Sec: discussion}

\begin{figure}%
\vspace{-1.3cm}
	\includegraphics[width=\columnwidth]{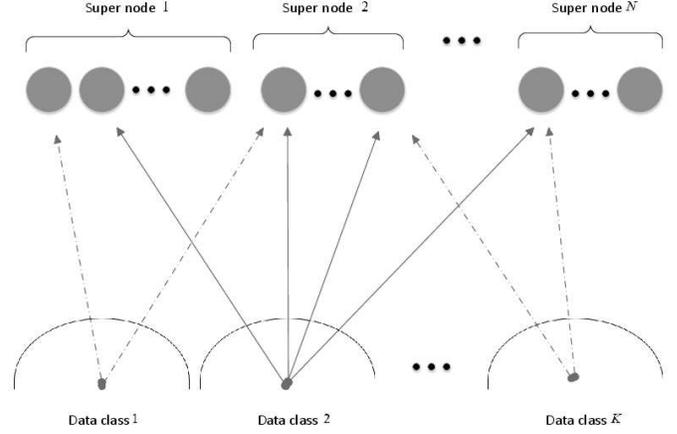}%
	\caption{An equivalent system model for the storage node capacity greater than one.}%
	\label{fig: Super node}%
\end{figure}

 Previously we assumed $c_n=1$ for all nodes. In other words, before normalizations, this is equivalent to say that each node can store the data of one class and that all nodes have the same capacity. Now, assume the storage capacity of nodes are different, also they can be greater than the size of data for one class. Normalizing these capacities with the data size of one class, node $n$ can store $c_n \ge 1$ unit(s) of data. Also, since our focus is on MSA, there is no point to consider non-integer $c_n$'s, hence $\forall n \in \mathcal{N} ,~ c_n \in \mathbb{N}$. 
 
 To handle such situations, we can think of node $n$ with capacity $c_n > 1$ as a super-node which consists of $c_n$ sub-storage nodes, each having unit capacity. This is depicted in Figure \ref{fig: Super node}. This new model is very similar to what we studied earlier where all nodes had unit capacity. The difference is in the access model. Here, we can think of two practically important cases.
 
 First, assume that the access to each sub-storage node is independent of all other sub-storage nodes, and the data collector successfully accesses each sub-storage node  with probability $p$. This can represent a practical scenario where different files are accessed in different times (e.g., requests are buffered at a storage node to be serviced later). In this case, the new model is equivalent to the MSA studied earlier and the optimal or efficient near-optimal solutions can be found using the previously discussed methods, by letting $N^{new}=\sum_{n=1}^{N} c_n$.

The other interesting case is when all the sub-storage nodes in one super node are accessed simultaneously by the data collector. This access is successful with probability $p$ and fails with probability $1-p$ for all these sub-storage nodes.  This for example can represent a situation when a hardware failure has affected a storage node (super-node).  For MSA, allocating more than one sub-storage node of a super-node to a specific data class is pointless. Thus, the total number of sub-storage nodes allocated to data class $i$ cannot be more than  the number of super nodes, i.e. $x_i \leq N  ,~  \forall i \in \mathcal{K}$. Moreover, as data are being assigned to sub-storage nodes, some super-nodes may ran out of capacity, putting an even harsher limit on the remaining data classes. 

To handle this case, we first solve MSA by letting  $N^{new}=\sum_{n = 1}^{N} c_n$ and $T_i^{new}= \min \{ T_i, |\mathcal{N}| \}  ,~ \forall i \in \mathcal{K}$. Assume $x_m^{opt}$ is the greatest among $x_i^{opt}$'s. Since we need to ensure that the total allocation given to class $m$ is more than that of class $j$ where $ j \in \mathcal{K} \setminus \{m \} $ , we start by allocating $x_m^{opt} \le N$ sub-storage units from $N$ different super-node. Since we do not want to fill super-nodes as long as possible, we start from the super-node with the largest capacity (super-node 1) and move on. After finishing allocation of $x_m^{opt}$, we update the available super-nodes since some may have been already filled. Assume these filled super-nodes are denoted by $\mathcal{N}_f$. We repeat solving MSA by letting  $\mathcal{N}^{new} = \mathcal{N} \setminus \mathcal{N}_f$, $\mathcal{K}^{new}  = \mathcal{K} \setminus \{m\}$. Note that since we allocate one data class in each iteration, this algorithm is $O(K (\sum_{n \in \mathcal{N}}^{} c_n))$. 

We have summarized the above procedure in Algorithm \ref{alg: optimal supernode}.

\begin{algorithm}[t]
	\caption{}\label{alg: optimal supernode}
	\begin{algorithmic}[1]
		\Procedure{}{$N, \mathcal{K},\cal{T}, \boldsymbol{\alpha}, c_n's$}

		\While {$K>0$}

		\State	$N \leftarrow \sum_{n \in \mathcal{N}}^{} c_n$ 
		
		\State	$T_i \leftarrow \min \{ T_i, |\mathcal{N}| \}  ,~ \forall i \in \mathcal{K}$
		
		\State $x_i \leftarrow 0  ,~ \forall i \in \mathcal{K} $
		
		\While {$N>0$}
		
		\State $\alpha_{j} \leftarrow max(\boldsymbol{\alpha}) $
		\State $ x_j \leftarrow x_j+1$
		\State $\alpha_j \leftarrow q \alpha_j$
		\State	$N \leftarrow N-1$

		\If	{$x_j \geq \floor{T_j} $ }		
		\State		$\boldsymbol{\alpha} \leftarrow ( \alpha_1, ..., \alpha_{I-1}, \alpha_{I+1}, ..., \alpha_K  ) $		
		\EndIf	
		
		\EndWhile

		\State $ x_m \leftarrow \max(x_i)$'s
		\State $ x^{opt}_m \leftarrow x_m$
		\State $\mathcal{N} \leftarrow \mathcal{N} \setminus \mathcal{N}_f$		
		\State $\mathcal{K}  \leftarrow \mathcal{K} \setminus \{m\}$

		\EndWhile
		
		\EndProcedure
	\end{algorithmic}
\end{algorithm}

\section{Simulation results} \label{Sec: simulation results}

In this section, simulation results are presented to verify our analytical analysis under different DSS setups.  

\begin{figure}%
	\includegraphics[width=1\columnwidth]{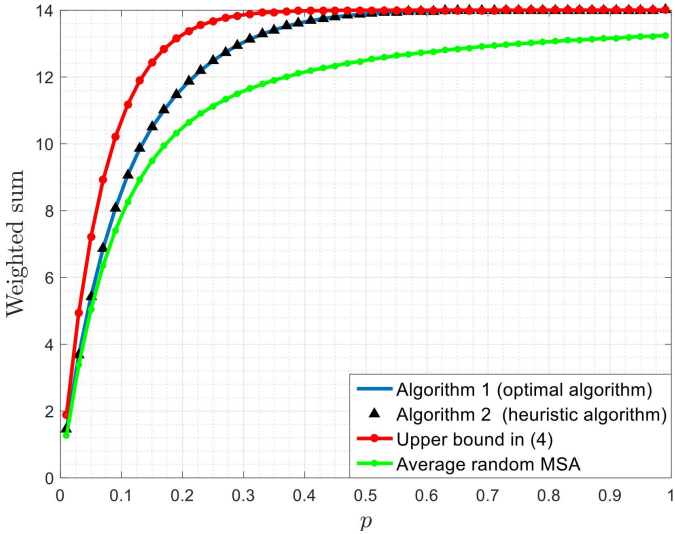}%
	\caption{Performance of optimal MSA in comparison with average random MSA.}%
	\label{fig: WeightedProbAccessBudget}%
\end{figure}

Figure~\ref{fig: WeightedProbAccessBudget} presents the simulations results for a multi-class DSS where $N=20$, $T_1=20$, $T_2=8$ and $T_3=4$. Further, the weights for each class are $\alpha_1=8$, $\alpha_2=5$, $\alpha_3=1$. More specifically, the weighted sums of the probabilities of successful recovery of all data classes for the optimal MSA obtained using Algorithm~\ref{alg:optimal} and the near-optimal MSA obtained using Algorithm~\ref{alg:WOMSA} are compared with the upper bound of (\ref{eq:bound}). As a benchmark, the weighted sum of the probabilities of successful recovery for a random MSA, averaged over 100 realizations, is also depicted in Figure~\ref{fig: WeightedProbAccessBudget}. As we see, our proposed solution in Algorithm~\ref{alg:WOMSA} matches the optimal solution of Algorithm~\ref{alg:optimal} and is in fact optimal. In addition, both solutions significantly outperform random MSA and achieve the upper bound for $p \geq 0.6$. This means that MSA is indeed the optimal allocation for \eqref{P1}.

\begin{figure}%
	\includegraphics[width=1\columnwidth]{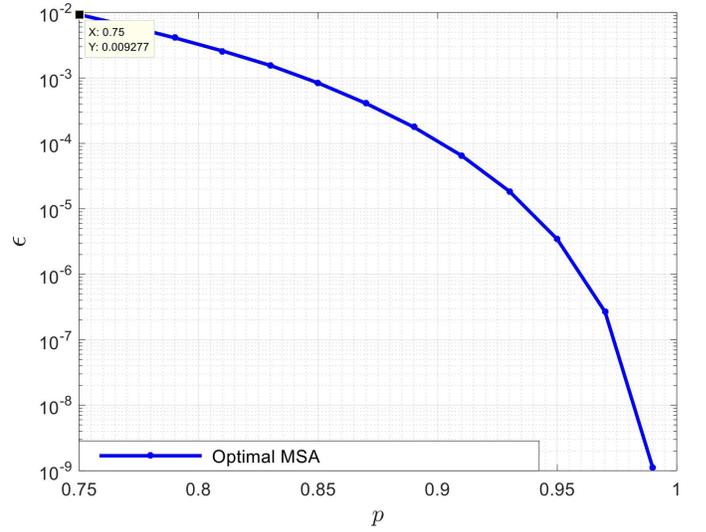}%
	\caption{The gap between the general upperbound and the optimal MSA.}%
	\label{fig: WeightedProbAccessWithoutBudget}%
\end{figure}

To verify our results in Theorem~\ref{Theorem: gap}, we present Figure~\ref{fig: WeightedProbAccessWithoutBudget} where three classes of data are stored over $N=15$ nodes. Here, it is assumed that $T1 = T_2 = T_3 = N$ and  $\alpha_1=6$, $\alpha_1=4$ and $\alpha_1=1$. The results of this figure confirm that the gap between the optimal MSA and the actual optimal storage allocation for \eqref{P1} approaches 0 and satisfies \eqref{eq theo5}.

\begin{figure}%
	\includegraphics[width=1\columnwidth]{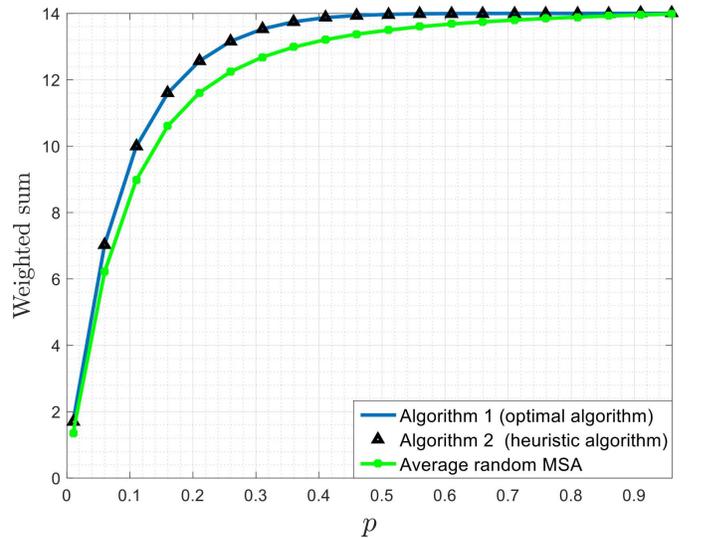}%
	\caption{Performance of the proposed solutions when at least one storage node has been dedicated to each data class.}%
	\label{fig: Remark1}%
\end{figure}

Figure \ref{fig: Remark1} depicts the results for a DSS with three classes of data where $N = 25$, $T_1=8$, $T_2=15$ and $T_3=23$. Further,  $\alpha_1=1$, $\alpha_1=5$ and $\alpha_1=8$ and it is assumed that $x_i^{\mathrm{min}} = 1$ for $i = 1, 2, 3$. Again, we observe that Algorithm~\ref{alg:WOMSA}'s results match the one for Algorithm~\ref{alg:optimal} verifying the (near) optimality of Algorithm~\ref{alg:WOMSA}. Further, both solutions outperform random MSA.


\section{Conclusion}

In this paper, we studied a multi-class DSS where various classes of data, each with a different QoS requirement in terms of successful recovery probability, are to be stored over the storage nodes. For access to the storage nodes, we considered the widely used probabilistic access model. We then formulated the optimization problem for finding the minimal spreading allocation (MSA) that maximizes the weighted sum of the successful recovery probability of all data classes, subject to a guaranteed probability of success for each class. Through a number of intermediate results we solved this optimization problem. We argued that the proposed solution is applicable to a wide range of scenarios including but not limited to heterogeneous DSSs with storage nodes that have unequal capacities. Next, we studied the gap between the performance of the optimal MSA allocation and some upper bounds on the optimal allocation. This study revealed that in many practical cases, MSA performs very close to the intractable optimal allocation. Finally, simulation results were presented to better illustrate our analysis.  

\section*{Acknowledgement}

The Authors would like to thank Alberta Innovates Technology Futures (AITF), TELUS Corporation and Natural Sciences and Engineering Research Council of Canada (NSERC) for supporting this work.

\appendices

\section{Proof of Theorem \ref{theorem state 2}}
\label{Appendix G}

\begin{IEEEproof}
	Assume $x_i = \floor{x_i}+ e_i, ~\forall i \in \mathcal{K}$ where $ 0 \leq e_i < 1  $. Without loss of generality, assume $\alpha_1 \geq \alpha_2 \geq ... \geq \alpha_K $ which implies that  $x_1\geq x_2 \geq ... \geq x_K $ (otherwise by changing the  $x_i$'s, one can decrease the objective function).

	Now note that we have $\sum_{i=1}^{K} x_i = N$ and $q^{x_i+ \log_q \alpha_i}=D, ~\forall i \in \mathcal{K}$ where $D$ is a constant and $D \in R^+$. 
	
Using contradiction, for the optimal MSA, assume $ \exists j  \in \mathcal{K},~ x^{opt}_j < \floor{x_j}  $, let $x^{opt}_j = \floor{x^{opt}_j}-h_j,~  h_j\geq 1$ which indicates $ \exists i\neq j \in \mathcal{K},~ x^{opt}_i > \floor{x_i}$ (since $\sum_{i=1}^{K} x^{opt}_i = N $) and let $x^{opt}_i = \floor{x_i} + h_i,~  h_i\geq 1$. Now, consider the new allocation as		
	\begin{numcases}{x'_k=}
	\floor{x_i} + h_i-1 & $k=  i$ \nonumber
	\\
	\floor{x_j}-h_j+1 & $k=  j$ \nonumber
	\\
	x^{opt}_k & $k \ne i,j$ \nonumber
	\end{numcases}		
	
	Notice that since we have $\mathcal{K}_3 = \emptyset$, the allocation strategy $x'_k$ is a feasible allocation. Now, we prove that the new allocation outperforms the former one. To this end, it is sufficient to only prove that

	\begin{eqnarray}\label{eqn: algo}
	\begin{aligned} 
	&                               &   & q^{\floor{x_i} + h_i +\log_q \alpha_i} + q^{\floor{x_j} - h_j +\log_q \alpha_j} \\
	&                               &   &\geq q^{\floor{x_i} + h_i-1 +\log_q \alpha_i} + q^{\floor{x_j} - h_j+1 +\log_q \alpha_j} \\
	&\stackrel {(a)}{\iff}          &   & D q^{h_i-e_i}  +  D q^{-h_j-e_j} \geq   q^{h_i-e_i-1} D +  q^{-h_j-e_j+1} D   \\
    &\stackrel {(b)}{\iff}          &   & q^{-h_j-e_j} (1-q) (1-q^{h_i+h_j+e_j-e_i-1})       \geq   0     \\
	\end{aligned}
	\end{eqnarray}
	where:\\
	($a$) follows from the fact that $q^{x_i+ \log_q \alpha_i}=D ,~\forall i \in \mathcal{K}$.\\
	($b$) follows from the facts that $e_j-e_i>-1$ and $ h_i+h_j \geq 2$.\\
	Equation \eqref{eqn: algo} contradicts the optimality of $x^{opt}_i$'s and shows that $x^{opt}_i \geq \floor{x_i} ,~ \forall i \in \mathcal{K}$. Similarly, it can be shown that $\nexists i \in \mathcal{K} ,~ x^{opt}_i > \floor{x_i}+1$.	
\end{IEEEproof}

%
%

\section{Proof of Theorem \ref{Theo state 2}}
\label{Appendix H}

\begin{IEEEproof}
		Note that since we have $\mathcal{K}_3 = \emptyset$,  $ x^{possible}_i = \floor{x_i}+1$ is a feasible allocation for $\forall i \in \mathcal{K}$.
	Using the proof of Theorem \ref{theorem state 2} and assuming $\exists  i,j  \in \mathcal{K} ,~ e_i \geq e_j  $, it is sufficient to show that	
	\begin{eqnarray} 
	\label{eqnnn}\nonumber
	&\alpha_i q^{\floor{x_i}+1} + \alpha_j q^{\floor{x_j} }  \leq \alpha_i q^{\floor{x_i}} + \alpha_j q^{\floor{x_j}+1 } & \\ \nonumber
	&\stackrel {(a)}{\iff}     q^{1-e_i} +  q^{-e_j }  \leq  q^{-e_i} + q^{1-e_j } & \\ \nonumber
	&\stackrel {(b)}{\iff}    (1-q)(q^{-e_j}-q^{-e_i}) \leq 0 & \nonumber
	\end{eqnarray} 	
	where:\\
	($a$) follows from the fact that $q^{x_i+ \log_q \alpha_i}=D ,~ \forall i \in \mathcal{K}$.\\
	($b$) follows from the fact that $e_i \geq e_j $.\\
	which completes the proof.

\end{IEEEproof}

\section{Proof of Theorem \ref{Theorem: gap}}
\label{Appendix J}

\begin{IEEEproof}	
	Using Theorem \ref{easy sol}, one can easily show that $x^{opt}_i \geq 1 ,~ \forall i \in \mathcal{K}$ if and only if $p \geq 1- (\frac{\alpha^{K-1}_{min}}{\prod_{j=1 , \alpha_j \neq \alpha_{min} }^{K}  \alpha_j} )^{ \frac{1}{|N-1|}  } $. Now, assume $x^{opt}_i \geq 1 , ~ \forall i \in \mathcal{K}$, for the performance gap between the perfect recovery (all data classes can be recovered with probability one) and optimal MSA we have  		
	\begin{equation*}
	\begin{aligned}
	& \sum_{i=1}^{K} \alpha_i -  \sum_{i=1}^{K} \alpha_i (1-q^{x^{opt}_i})= \sum_{i=1}^{K}  q^{x^{opt}_i+\log_q \alpha_i   }  &    \\
	& \geq K q^{\frac{\sum_{i=1}^{K} x^{opt}_i +  \log_q \prod_{i=1}^{K} \alpha_i  }{K}}  =  K q^{\frac{N +  \log_q \prod_{i=1}^{K} \alpha_i  }{K}}  \geq \epsilon                         &	   
	\end{aligned}
	\end{equation*}

	\vspace{0.5em}	
	
	\noindent the first inequality comes from the geometric mean arithmetic-mean inequality.	
	Solving the last inequality for $p=1-q$ gives $p \leq 1- (\frac{\epsilon^K}{K^K \prod_{i=1}^{K} \alpha_i})^{\frac{1}{N}}$ which further implies that the performance gap between the optimal MSA and perfect recovery is less than $\epsilon$ for
	$\displaystyle{p> \max \{1- (\frac{\epsilon^K}{K^K \prod_{i=1}^{K} \alpha_i})^{\frac{1}{N}} ,  1- (\frac{\alpha^{K-1}_{min}}{\prod_{j=1 , \alpha_j \neq \alpha_{min} }^{K}  \alpha_j} )^{|\frac{N}{K}-1|} }  \}$ and  completes the proof.	
	
\end{IEEEproof}

\bibliographystyle{IEEEtran}
\bibliography{IEEEabrv,reference}

\begin{thebibliography}{10}
\providecommand{\url}[1]{#1}
\csname url@samestyle\endcsname
\providecommand{\newblock}{\relax}
\providecommand{\bibinfo}[2]{#2}
\providecommand{\BIBentrySTDinterwordspacing}{\spaceskip=0pt\relax}
\providecommand{\BIBentryALTinterwordstretchfactor}{4}
\providecommand{\BIBentryALTinterwordspacing}{\spaceskip=\fontdimen2\font plus
\BIBentryALTinterwordstretchfactor\fontdimen3\font minus
  \fontdimen4\font\relax}
\providecommand{\BIBforeignlanguage}[2]{{%
\expandafter\ifx\csname l@#1\endcsname\relax
\typeout{** WARNING: IEEEtran.bst: No hyphenation pattern has been}%
\typeout{** loaded for the language `#1'. Using the pattern for}%
\typeout{** the default language instead.}%
\else
\language=\csname l@#1\endcsname
\fi
#2}}
\providecommand{\BIBdecl}{\relax}
\BIBdecl

\bibitem{leong2012distributed}
D.~Leong, A.~G. Dimakis, and T.~Ho, ``Distributed storage allocations,''
  \emph{{IEEE} Trans. Inf. Theory}, vol.~58, no.~7, pp. 4733--4752, 2012.

\bibitem{Sardari_Allocation_2010}
M.~Sardari, R.~Restrepo, F.~Fekri, and E.~Soljanin, ``Memory allocation in
  distributed storage networks,'' in \emph{IEEE Intl. Symp. on Information
  Theory (ISIT)}, June 2010, pp. 1958--1962.

\bibitem{ntranos2012allocations}
V.~Ntranos, G.~Caire, and A.~G. Dimakis, ``Allocations for heterogenous
  distributed storage,'' in \emph{IEEE Intl. Symp. on Information Theory
  (ISIT)}, 2012, pp. 2761--2765.

\bibitem{hong2016asymptotic}
B.~Hong and W.~Choi, ``Optimal storage allocation for wireless cloud caching
  systems with a limited sum storage capacity,'' \emph{{IEEE} Trans. Wireless
  Commun.}, vol.~15, no.~9, pp. 6010--6021, Sept 2016.

\bibitem{Li2013Allocation}
Z.~Li, T.~Ho, D.~Leong, and H.~Yao, ``Distributed storage allocation for
  heterogeneous systems,'' in \emph{Allerton Conf. on Communication, Control,
  and Computing}, Oct 2013, pp. 320--326.

\bibitem{Noori2015k-guaranteed}
M.~Noori and M.~Ardakani, ``Allocation for heterogeneous storage nodes,''
  \emph{IEEE Communications Letters}, vol.~19, no.~12, pp. 2102--2105, Dec
  2015.

\bibitem{andriyanova2015distributed}
I.~Andriyanova and P.~M. Olmos, ``On distributed storage allocations for
  memory-limited systems,'' in \emph{IEEE Global Communications Conf.
  (GLOBECOM)}, 2015, pp. 1--6.

\bibitem{LeongOptimaldelay2011}
D.~Leong, A.~G. Dimakis, and T.~Ho, ``Distributed storage allocations for
  optimal delay,'' in \emph{IEEE Intl. Symp. on Information Theory (ISIT)},
  July 2011, pp. 1447--1451.

\bibitem{joshi2014delay}
G.~Joshi, Y.~Liu, and E.~Soljanin, ``On the delay-storage trade-off in content
  download from coded distributed storage systems,'' vol.~32, no.~5, pp.
  989--997, 2014.

\bibitem{Noori16Servicerate}
M.~Noori, E.~Soljanin, and M.~Ardakani, ``On storage allocation for maximum
  service rate in distributed storage systems,'' in \emph{IEEE Intl. Symp. on
  Information Theory (ISIT)}, July 2016, pp. 240--244.

\bibitem{yu2011minimization}
Q.~Yu, K.~W. Shum, and C.~W. Sung, ``Minimization of storage cost in
  distributed storage systems with repair consideration,'' in \emph{IEEE Global
  Communications Conf. (GLOBECOM)}, 2011, pp. 1--5.

\bibitem{Kumar_Multi_2015}
A.~Kumar, R.~Tandon, and T.~Clancy, ``On the latency and energy efficiency of
  distributed storage systems,'' \emph{{IEEE} Trans. Cloud Computing}, vol.~PP,
  no.~99, pp. 1--13, 2015.

\bibitem{Amazon}
\url{https://aws.amazon.com/s3/storage-classes/}, [Online; accessed
  22-Nov-2016].

\end{thebibliography}

\end{document}